\documentclass[12pt]{article}
\usepackage{amsmath,amssymb,epsfig}

\usepackage{color}
\input{colordvi.tex}

\newcommand{\lsim}{\lesssim}
\newcommand{\gsim}{\gtrsim}

\makeatletter

\renewcommand\section{\@startsection {section}{1}{\z@}%
                                   {-3.5ex \@plus -1ex \@minus -.2ex}%
                                   {2.3ex \@plus.2ex}%
                                   {\normalfont\large\bfseries}}

\renewcommand\subsection{\@startsection{subsection}{2}{\z@}%
                                     {-3.25ex\@plus -1ex \@minus -.2ex}%
                                     {1.5ex \@plus .2ex}%
                                     {\normalfont\normalsize\bfseries}}

\newcount\hour \newcount\minute
\hour=\time \divide \hour by 60
\minute=\time
\def\now{%
\ifnum \hour<13
  \ifnum \hour=0 \advance \hour by 12 \number\hour:\else \number\hour:\fi%
     \ifnum \minute<10 0\fi%
     \number\minute%
\ A.M.%
\else \advance \hour by -12 \number\hour:%
  \ifnum \minute<10 0\fi%
  \number\minute%
  \ P.M.%
\fi%
}

\makeatother

\begin{document}

\baselineskip=18pt  
\numberwithin{equation}{section}  
\allowdisplaybreaks  



%
%


\thispagestyle{empty}

\vspace*{-2cm}
\begin{flushright}
\end{flushright}

\begin{flushright}
MCTP-10-12\\
NSF-KITP-10-032\\
\end{flushright}

\begin{center}

\vspace{2.4cm}

{\bf\Large Light Gauginos and Conformal Sequestering}
\\

\vspace*{1.5cm}
{\bf
Kentaro Hanaki$^{1}$, and Yutaka Ookouchi$^{2,3}$} \\
\vspace*{0.5cm}

$^{1}${\it {Michigan Center for Theoretical Physics, University of Michigan, MI, 48109 USA}}
\vspace{0.1cm}

$^{2}${\it Perimeter Institute for Theoretical Physics, Ontario N2L2Y5, Canada}\\
\vspace{0.1cm}

$^{3}${\it {Kavli Institute for Theoretical Physics, University of California, CA, 93106, USA}}

\vspace*{0.5cm}

\end{center}

\vspace{1cm} \centerline{\bf Abstract} \vspace*{0.5cm}

In a wide class of direct and semi-direct gauge mediation models, it has been observed that the gaugino masses vanish at leading order. It implies that there is a hierarchy between the gaugino and sfermion masses, invoking a fine-tuning problem in the Higgs sector via radiative corrections. In this paper, we explore the possibility of solving this anomalously light gaugino problem exploiting strong conformal dynamics in the hidden sector. With a mild assumption on the anomalous dimensions of the hidden sector operators, we show that the next to leading order contributions to the gaugino masses can naturally be in the same order as the sfermion masses. $\mu/B_\mu$ problem is also discussed.

\newpage
\setcounter{page}{1} 





\section{Introduction}

Gauge mediated supersymmetry breaking \cite{GMI,GMII,GMIII,GGM} is a highly predictive and attractive way of transmitting supersymmetry breaking of the hidden sector to the SSM (See \cite{Giudice:1998bp,Intriligator:2007cp,Kitano:2010fa} for reviews). Also, given the fact that gauge interactions do not distinguish flavors, phenomenologically dangerous flavor changing neutral currents are naturally suppressed.

One drawback of gauge mediation is that gauginos tend to be light compared to sfermions. First of all, 
gauge interactions themselves do not break R-symmetry, so R-symmetry must be broken in the hidden sector for gauginos to obtain non-vanishing masses. But as was studied in \cite{NS}, the absence of supersymmetry preserving vacuum for a theory with generic superpotential requires R-symmetry be spontaneously broken, thus introducing a phenomenologically unfavorable R-axion. This problem can be avoided using recently proposed metastable supersymmetry breaking vacua \cite{ISS}. However, there is another problem: broken R-symmetry itself does not guarantee sufficiently large gaugino masses. Actually, as initially pointed out in \cite{IzawaTobe} for direct gauge mediation and in \cite{SeqI} for semi-direct gauge mediation, in a wide class of direct \cite{DM,Terning,DirectAbel,Maru,Strassler} and semi-direct \cite{Seiberg:2008qj,Elvang:2009gk,Argurio} gauge mediation models, the leading order contribution to gaugino masses vanishes irrespective of whether R-symmetry is broken or not\footnote{For $F$-term breaking models with calculable messenger sector, one can show that there is an upper bound for the ratio of gaugino masses to sfermion masses \cite{GMGM}.}. 


Anomalously light gauginos are problematic because relatively heavy sfermions induce a large correction to the Higgs mass, reintroducing the hierarchy problem. One possible way out may be to take the messenger scale to be very close to the supersymmetry breaking scale so that the subleading corrections are to be in the same size as the leading contribution. However, as was studied in \cite{Yonekura}, such a model is severely constrained by the recent Tevatron bound on the sparticle masses and the mass bound on a light gravitino.

Recently, Komargodski and Shih shed light on the origin of the light gauginos. In \cite{KS}, they related the vanishing gaugino masses at leading order and global structure of the vacua in renormalizable theories, and showed, based on the study of generalized O'Raifeartaigh models, that the pseudomoduli space must have a tachyonic direction somewhere to generate sizable gaugino masses. This analysis opens up a new possibility to avoid the anomalously light gaugino problem. Namely, the leading order gaugino mass generally does not vanish if  supersymmetry is broken in uplifted metastable vacua. This idea was initially employed in \cite{KOO} and further discussed in \cite{GKK,AbelKhoze,Mc}\footnote{For recent studies for generating leading order gaugino masses in semi-direct gauge mediation, see \cite{IIN1}.}.

In this paper, we propose an alternative solution to this anomalously light gaugino mass problem. 
As was studied in \cite{AGLR}, due to the gaugino screening effect, it is hard to generate the leading order gaugino mass at one-loop level even if we impose a non-canonical K\"ahler potential of a messenger. Thus, we instead reduce the sfermion masses relative to the gaugino masses by exploiting conformal sequestering \cite{Luty:2001jh,Luty:2001zv}\footnote{For applications to gauge mediation, see, e.g., \cite{SeqI,SeqII,SeqIII,SeqIV,SeqV,SeqVI}.}: With an appropriate choice of a hidden sector which is assumed to approach a strongly coupled fixed point below the messenger scale, sfermion masses can be suppressed relative to the gaugino masses by strong hidden sector renormalization group effects. Thus, one can make the sfermion mass parameters be lower or in the same order as the gaugino masses at the end of conformal sequestering. Then, the sfermion masses are driven to be in the same order as gaugino masses by standard model renormalization group effects.

In this work, we only discuss how conformal sequestering works for a messenger model with vanishing leading order gaugino masses and do not discuss the details of the hidden sector. But embedding our scenario into direct and semi-direct gauge mediations where the gaugino masses are zero at leading order would be possible and an interesting future direction.

The rest of the paper is organized as follows. In section 2, we review light gaugino mass problem and gaugino screening. Then, we comment on UV sensitivity of a non-renormalizable model and discuss a contribution of a heavy messenger which is either coupled to a light messenger or not. In section 3, we discuss the effects of conformal sequestering on sfermion and gaugino masses as well as $\mu$, $B_{\mu}$ terms. By showing explicit mass scales, we demonstrate that our scenario can be phenomenologically viable. In the Appendix, we present a model of Higgs-messenger coupling which generates $\mu$ and $B_\mu$ at one-loop with a messenger sector exhibiting vanishing leading order gaugino masses.

\section{Anomalously Light Gaugino Problem}

In this section, we review some known facts on gaugino masses at leading order in supersymmetry breaking scale. If the SUSY breaking scale is much smaller than messenger masses, one can reliably use the technique of analytic continuation into superspace \cite{GR,AGLR}.

\subsection{Leading order gaugino mass}

We first review anomalously small gaugino mass problem which have been observed quite frequently in direct gauge mediation models. Suppose a messenger sector having the following general superpotential interaction with supersymmetry breaking field $\langle X \rangle =M + \theta^2 F$,
\begin{eqnarray}
W=\sum_{ab} {\cal M}(X)_{ab} \phi^a \tilde{\phi}^b .
\label{Messenger}
\end{eqnarray}
where the messenger mass matrix $\mathcal{M}_{ab}(X)$ is a holomorphic function of $X$. In this case, the gaugino masses are generated by integrating out the messengers $\phi^a$, $\tilde{\phi}^b$. The formula is given \cite{GR,KOO} by
\begin{equation}
m_{ \lambda}=-{g^2_{\rm SM} \over 16 \pi^2} F{\partial \over \partial X} \log \det \mathcal{M}(X).
\label{GauginoMass}
\end{equation}
As was studied in \cite{KS}, if the fermion mass matrix ${\cal M}(X)$ includes a zero eigenvalue at some point in the pseudomoduli space spanned by $X$, then the corresponding bosonic mode becomes tachyonic in a region including the point where the fermionic zero mode appears. In this case, the determinant of ${\cal M}(X)$ is a function of $X$. On the other hand, if the eigenvalues of ${\cal M}(X)$ are non-vanishing, the bosonic modes of $\phi^a$ are stable everywhere\footnote{The presence of a supersymmetric vacuum at large $X$ should be a small effect. } in the pseudomoduli space and $\det {\cal M}(X)$ is constant, which yields vanishing leading order gaugino masses. Since the sfermion masses are generally generated at this order, the gauginos are light relative to the sfermions if the pseudomoduli space is stable everywhere.

\subsection{Gaugino screening}

Now we consider a more general setup in which the K\"ahler potential of messengers is non-canonical: it has a dependence on the supersymmetry breaking field $X$. Let us derive the gaugino mass formula utilizing analytic continuation into superspace \cite{AGLR}.  Suppose the theory has $N$ pair of messengers $\phi^a, \tilde{\phi}^a \,\, (a=1, \dots, N)$ which are fundamentals and anti-fundamentals of the standard model gauge interactions, respectively. A generic K\"ahler potential we consider is
\begin{equation}
K=\sum_a {Z}_a(X,X^{\dagger})(\phi^{a \dagger} e^{V_{\rm SM}^{(\phi)}}\phi^a + \tilde{\phi}^{a \dagger} e^{V_{\rm SM}^{(\tilde{\phi})}}\tilde{\phi}^a ), \label{KahlerA}
\end{equation}
where ${Z}_a(X,X^{\dagger})$ are some real functions of $X, X^{\dagger}$.
Finally, the superpotential is given by \eqref{Messenger}.

One can extract the gaugino masses generated by integrating out the messenger fields from the wave function renormalization gauge chiral superfield. One should, however, use the physical gauge coupling $R$ rather than the holomorphic one, since the holomorphic coupling is not invariant under field rescaling \cite{AGLR}. As pointed out in \cite{AGLR}, contributions from messenger interactions to the gaugino masses are suppressed by additional loop factors. Thus, a non-canonical K\"ahler potential cannot contribute to the leading order gaugino mass. To see this, one may write down the physical coupling below the messenger scale. For the sake of simplicity, we assume the fermion mass matrix of the messengers is constant: ${\cal M}(X)=m$, so $W=m \phi \tilde{\phi}$. The physical mass is defined using wavefunction renormalization ${ Z}_M$ of the messenger at the scale,
$${
\mu_m^2 ={|m|^2 \over { Z}_M(\mu_m)^2}.
}$$
Below this scale, the physical coupling is given by
$${
R(\mu)=R^{\prime}(\mu_0)+{b \over 16\pi^2} \log {\mu^2 \over \mu_0^2}+ {1 \over 16 \pi^2} \log {|m|^2 \over \mu_0^2 { Z}^{\prime}_M (\mu_0)^2}+{T_G \over 8\pi^2}\log {{\rm Re} S(\mu)\over {\rm Re} S^{\prime}(\mu_0)}-\sum_r {T_r \over 8\pi^2} \log { { Z}_r (\mu)\over { Z}_r^{\prime}(\mu_0) },
}$$
where $r$ runs all matter fields in the SSM, $\mu_0$ is the cut-off scale of the theory, and $b$ is a coefficient of beta function below the messenger scale. $S(\mu)$ is a holomorphic gauge coupling and primed quantities are the ones above the messenger scale. Here, we see that ${{ Z}_M(\mu_m)}$ dependence drops out at low energy. Thus, a non-canonical K\"ahler potential does not contribute to the leading order of gaugino mass. Moreover, we could have assumed a spurion dependence of K\"ahler potential at the cut-off scale. Plugging the definition of real coupling $R^{\prime}(\mu_0)$ at the cut-off scale, we see that ${ Z}^{\prime}_{M}(\mu_0)$ dependence also cancels out. Therefore, the leading order gaugino masses are not affected by spurion dependence of K\"ahler potential of messengers at all. However, if we impose a spurion dependence in $S^{\prime}(\mu_0)$, it definitely contributes. Although it is nothing but adding gaugino masses by hand at the cut-off scale, it is contained in a frame work of the gauge mediation \cite{GGM}, since it vanishes in turning off the gauge coupling of the SSM. Usually in calculable models, these contributions, if exist, are generated by a heavy messenger around the cut-off scale and small compared to the leading term. Nevertheless, as we will see below, in some case it can be dominant and comparable to the sfermion masses.

\subsection{Next to leading order gaugino mass}

Since the sfermion masses generally arise at leading order, the vanishing gaugino masses at leading order implies that there is a hierarchy between gaugino and sfermion masses. One may consider the next to leading order gaugino masses to solve the hierarchy. There are several sources for non-vanishing gaugino masses at next to leading order\footnote{Here, we focus on the gaugino masses generated at or above the messenger scale. At low-energy, the Higgs-Higgsino loop can also generate non-zero gaugino masses \cite{SYY}.}:
\begin{enumerate}

\item While the gaugino masses leading order in $F$ at one-loop are prohibited, there is no problem for having non-vanishing gaugino masses at higher order in $F$. Explicit calculations show that the next leading order contribution arises at ${\cal O} ({F^3/ M_{\rm mess}^5})$ \cite{IzawaTobe}. One might hope that the gaugino masses can be comparable to sfermion masses if $F/M_{\rm mess}^2 \sim 1$. However, these higher order corrections are suppressed by small numerical coefficients in known examples and not sufficient to solve the hierarchy. Also, there is a phenomenological constraint on such a low scale mediation model \cite{Yonekura}.

\item Another possibility is that the gaugino masses are generated in ${\cal O}(F)$, but at higher loop level. In this case, using wave-function renormalization technique \cite{AGLR}, one can explicitly show that the leading order gaugino mass in $F$ at two-loop also vanishes if one-loop contribution does. Thus, the leading contribution is generated at best from three-loop diagrams. This contribution includes additional loop factors, so should be suppressed compared to the leading order. 

\item As discussed in the previous subsection, there could be a contribution of a heavy messenger at the cut off scale or above, which would be of order ${\cal O}(F/M_{\rm heavy})$. This type of contribution is suppressed to the leading order soft masses by a factor of ${\mathcal O} (M_{\rm mess}/M_{\rm heavy})$ compared to the leading order soft masses.

\end{enumerate}
In any case, the gaugino masses are suppressed compared to the leading order and so to sfermion masses. This, combined with the current experimental lower bound for gaugino masses, indicates that the scales of the sfermion masses should be much higher than that of electroweak symmetry breaking, giving rise to fine tuning for the Higgs mass via top-stop loops. This is in contrast to the fact that relatively heavy gauginos at the messenger scale does not cause any problem because the sfermion masses are driven to be in the same order as gaugino masses at a lower scale by standard model renormalization group effects. In the next section, we explore the possibilities of solving this fine tuning problem using strong conformal dynamics in the hidden sector.
 
\section{Conformal Sequestering}

In the previous section, we learned that the vanishing of the gaugino masses at leading order gives rise to fine tuning associated to heavy sfermions and just modifying the K\"ahler potential does not generates a large enough gaugino masses. In this section, we show that the gaugino masses can naturally be larger than or comparable to the gaugino masses if the hidden sector flows into a strongly coupled fixed point below the messenger scale. We first see how the strong conformal dynamics in the hidden sector reduces the sfermion masses. Then we show how the next to leading order gaugino masses are sequestered, and provide a set of constraints the parameters should satisfy to avoid fine tuning. As the hidden sector is assumed to be strongly coupled, one cannot calculate the anomalous dimensions of the non-holomorphic operators. But we expect that given a large number of SCFTs, one can find appropriate SCFTs which satisfy the constraints. To illustrate, we discuss the explicit numbers which are consistent with the constraints. We end this section with a discussion on $\mu/B_{\mu}$ problem. We assume that supersymmetry is broken by a gauge singlet throughout this section.

\subsection{Sequestering of scalar masses}

\begin{figure}[tbp]
\begin{center}
\includegraphics[width=4.5cm,height=6.0cm]
{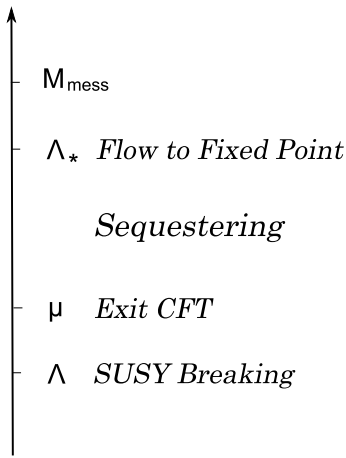}
\end{center}
    \caption{\small \it Energy scales in our scenario}
\end{figure}

Let us assume that a singlet chiral superfield $X$ eventually gets a non-zero $F$-component vev and breaks SUSY. Before the hidden sector flows into the IR fixed point, messengers are integrated out at the scale $M_{\rm mess}$. It generates interactions between $X$, $X^\dagger$ and the SSM matters in K\"ahler potential and subsequently the sfermion masses. The operators which generate the sfermion masses are summarized as\footnote{We impose messenger parity \cite{GMI,Mparity} to prevent dangerous hypercharge $D$-terms from being generated in the hidden sector.}
$${
\int d^4 \theta\,  T(X,X^{\dagger})\,  Q^{\dagger}_{\rm SM}Q_{\rm SM}.
}$$
and the $D$-component of $T(X,X^\dagger)$ induces the sfermion masses. In general, the function 
$T(X,X^\dagger)$ can be expanded as follows:
$${
T(X,X^\dagger)=c_0 |X|^2+\cdots
}$$
where the ellipsis contains a holomorphic and anti-holomorphic function of $X$ which do not contribute to the scalar masses as well as higher order interaction terms.

Our goal is to suppress these terms so that the sfermion masses can be lower than or in the same order as next to leading order gaugino masses. To do this, let us suppose that at some scale $\Lambda_*$ below $M_{\rm mess}$, the theory flows into a strongly coupled CFT (see Figure $1$). Since the theory is strongly coupled, the anomalous dimensions of various operators such as $X$ and $|X|^2$ are expected to be order one\footnote{Since the conserved current is not renormalized, one have to be careful about global symmetries which may prevent these operators from acquiring large anomalous dimensions \cite{Luty:2001jh,Luty:2001zv}.}. Then, the terms in the K\"ahler potential, or equivalently those in $T(X)$, are suppressed until the theory exits from the conformal regime at a scale $\mu$. If the anomalous dimensions of $X$ and $|X|^2$ are given by $\gamma_X$ and $2 \gamma_X + \alpha_*$, respectively, the operators in $T(X,X^\dagger)$ are sequestered as
\begin{eqnarray}
T(X,X^\dagger) \to \epsilon^{2\gamma_X+\alpha_*} c_0 |X|^2 +\cdots,
\end{eqnarray}
where the epsilon is defined by
$${\epsilon=\mu/\Lambda_*.
}$$
Note that the anomalous dimension of ${|X|^2}$ is not just the twice of the anomalous dimension $\gamma_{X}$ because it is not a chiral operator and this fact is crucial to suppress the sfermion masses relative to the gaugino masses. Then, after the theory exits from a strongly coupled CFT at $\mu$, supersymmetry is broken at the scale $\Lambda^2 :=F < \mu^2$.

In this scenario, the leading order of the scalar masses below the scale $\mu$ are given by
\begin{equation}
m^2_{\rm s}= \epsilon^{2\gamma_X + \alpha_*}\, \left(\frac{g^2_{\rm SM}}{16 \pi^2}\right)^2 c_0 |F|^2 \sim \epsilon^{2\gamma_X + \alpha_*}\, \left(\frac{g^2_{\rm SM}}{16 \pi^2}\right)^2 { |F|^2 \over M_{\rm mess}^2} ,
\end{equation}
where we took $c_0 \simeq 1/|M_{\rm mess}|^2$ because this operator was generated by integrating out the messengers with mass $M_{\rm mess}$\footnote{For simplicity, we also assumed that higher order terms have larger anomalous dimensions and so suppressed by larger powers of $\epsilon$. Since the hidden sector is strongly coupled, however, there is no way to justify this assumption.}.

Let us then discuss the sequestering of the gaugino masses. As discussed in the previous section, the following three contributions to the gaugino masses can be dominant when the leading order contribution \eqref{GauginoMass} vanishes:

\begin{enumerate}

\item The first possibility is the case where the gaugino masses generated at order $F^3/M_{\rm mess}^5$ at one-loop level dominate over others. These gaugino masses typically originate from non-holomorphic operators including $D$ or $\bar{D}$. For example, a term
\begin{equation}
\int d^4 \theta \frac{X^\dagger XX^\dagger D^2 X}{M_{\rm mess}^6} W^\alpha W_\alpha,  \nonumber
\end{equation}
generates gaugino masses of order $F^3/M_{\rm mess}^5$ if one assumes that the lowest component of $X$ is in the same order as $M_{\rm mess}$. This type of operators is not protected by the superconformal algebra, so one cannot determine the precise scaling at strong coupling. Nevertheless, one expects that for an appropriate choice of SCFT, the gaugino masses can be at the same order as the sfermion masses after sequestering. Let us assume that the anomalous dimension of the operator generating the ${\mathcal O}(F^3/M_{mess}^5)$ correction to be $4 \gamma_X + \gamma_*$. Then, the gaugino masses after conformal sequestering are
$${
m^{\rm 1-loop}_{\lambda} \sim \epsilon^{4 \gamma_X+\gamma_*} \frac{g^2_{\rm SM}}{16 \pi^2}{F^3\over M_{\rm mess}^5} > m^{\rm 3-loop}_{\lambda}, m_{\lambda}^{\rm heavy}.
}$$
If the condition $\epsilon^{\alpha_*/2 -3 \gamma_X - \gamma_*}\simeq F^2/M_{\rm mess}^4$ holds, the gaugino mass is comparable to the sfermion masses.

\item The second possibility comes from the leading contribution in $F$ at three loop level. As noted in the previous section, at two loop order the leading contribution in $F$ also vanishes. When this contribution dominates over others, 
$${
m^{\rm 3-loop}_{\lambda} \sim \epsilon^{\gamma_X} \left(\frac{g^2_{\rm SM}}{16 \pi^2}\right)^3 {F\over M_{\rm mess}} > m^{\rm 1-loop}_{\lambda}, m_{\lambda}^{\rm heavy},
}$$
when $\epsilon^{\alpha_*/2} \sim (g^2_{\rm SM}/16 \pi^2)^2$, the gaugino and scalar masses are in  the same order. This scenario, however, is not desirable. The gaugino masses at three-loop level is proportional to $\alpha_{\rm SM}^3$. Since $g_3^2/g_1^2 \sim 7$ at TeV scale, there is a ${\mathcal O} (100-1000)$ hierarchy between the bino and gluino masses. Since the squarks are charged under $SU(3)$ gauge group in the standard model, the heavy gluino mass induces a large standard model renormalization group effects on squark masses, making the squark masses be roughly at the same order as gluino mass, invoking a fine tuning problem. Hence, we do not consider this possibility in the rest of the paper.

\item The last possibility is coming from a decoupled heavy messenger around the cut-off scale. There could be a contribution to the gaugino mass at leading order in $F$ at one loop level. Since it typically comes from a operator $XW_{\alpha}W^{\alpha}$ generated at the heavy messenger scale $M_{\rm heavy}$, one expect the following scaling under the sequestering.  
$${
m^{\rm heavy}_{\lambda} \sim \epsilon^{\gamma_X} \frac{g^2_{\rm SM}}{16 \pi^2} {F\over M_{\rm heavy}} > m^{\rm 1-loop}_{\lambda}, m_{\lambda}^{\rm 3-loop}.
}$$
When this contribution dominates over others, by taking $\epsilon^{\alpha_*/2}\simeq M_{\rm mess}/M_{\rm heavy}$ one can make gaugino mass and scalar mass are in the same order. 

\end{enumerate}
Which effect can be dominant is the subject of the next subsection.

\subsection{Constraints on the parameters}

For simplicity, let us assume that the hidden sector enters into the conformal regime somewhere close to the messenger scale $\Lambda_* \sim M_{\rm mess}$ and it exits near the scale of supersymmetry breaking, $\mu \sim \sqrt{F}$. For the first scenario, where one-loop $F^3$ contribution is dominant, to be realized, three-loop contribution should be sufficiently suppressed. This condition leads to
\begin{equation}
\epsilon^{3 \gamma_X + \gamma_* + 4} > \left( \frac{\alpha_{\rm SM}}{4 \pi}\right)^2.
\label{F3Dominant}
\end{equation}
This inequality implies that the lower bound of $\epsilon$ becomes larger for larger $3 \gamma_X + \gamma_*$. Especially, if $3 \gamma_X + \gamma_*$ is positive, $\epsilon$ must be larger than ${\mathcal O}(10^{-1})$. If we assume that the coupling in the hidden sector is small at the messenger scale, this may not be enough to guarantee the room for the small coupling constant at the messenger scale to become large at the conformal fixed point, so the conformal sequestering may not happen. For strongly coupled messengers, however, this is not a problem. Another constraint is that the gaugino masses should be comparable to or larger than the sfermion masses. This implies
\begin{equation}
\epsilon^{\alpha_*/2 - 3 \gamma_X - \gamma_* - 4} \lsim 1.
\label{F3Comparable}
\end{equation}
This condition may require large $|\alpha_*|$ and $|\gamma_*|$ while they can still be order one. Since $\epsilon<1$, one can rewrite the inequality \eqref{F3Comparable} as
\begin{equation}
\frac{\alpha_*}{2} - \gamma_* \gsim 4 + 3 \gamma_X.
\end{equation}
In any CFT, the conformal dimension of a gauge singlet, Lorentz scalar operator must be no less than one, thus $\gamma_X$ must be positive. For $\gamma_X \gsim 2$, the left hand side of \eqref{F3Comparable} should be larger than 10, so it requires somewhat large $|\alpha_*|$ and $|\gamma_*|$, while for small $\gamma_X$, $|\alpha_*|$ and $|\gamma_*|$ can be order one. Finally, we take the mass of the lightest gaugino, bino in this case, to be of order $10^2$ GeV to avoid fine tuning:
\begin{equation}
\frac{g_{1}^2}{16 \pi^2}M_{\rm mess} \epsilon^{4 \gamma_X + \gamma_* + 6} \sim 10^2\; {\rm GeV},
\label{F3SfermionMasses}
\end{equation}
which determines the size of $M_{\rm mess}$.

The argument above may be too abstract, so we can use numerical values to see that this scenario can be realized with order one anomalous dimensions. Let us take
\begin{equation}
\gamma_X \sim 0.5, \quad \gamma_* \sim -4, \quad \epsilon \sim 10^{-1.5}.\nonumber
\end{equation}
Then, \eqref{F3Dominant} is satisfied. Eq.\eqref{F3Comparable} determines the value of $\alpha_*$ to be $\gsim 3$, so let us just take $\alpha_* \sim 3$. The messenger scale and the supersymmetry breaking scale are fixed by \eqref{F3SfermionMasses} and 
\begin{equation}
M_{\rm mess} \sim 10^{11}\;{\rm GeV}, \quad \sqrt{F} \sim 10^{9.5}\;{\rm GeV}.  \nonumber
\end{equation}
For this choice of parameters, we can also calculate the gravitino mass
\begin{equation}
m_{3/2} = \frac{F}{M_{\rm Pl}} \sim {\mathcal O}(1 \; {\rm GeV}).  \nonumber
\end{equation}
So, the anomaly and gravity mediation effects are also suppressed compared to the gauge mediation contributions. Note that the gravity mediation effect is also suppressed by conformal sequestering.

For the third scenario to be realized, the effect of heavy messengers should dominate over the other contributions. One-loop, $F^3$ contribution is naturally small if $\gamma_*$ is large. The condition that the contribution from heavy messengers is larger than that from three-loop contribution gives
\begin{equation}
\frac{M_{\rm mess}}{M_{\rm heavy}} > \left( \frac{g_{\rm SM}^2}{16 \pi^2} \right)^2.
\end{equation}
For the light messengers to play some role, the gaugino masses from the heavy messenger should be comparable to the sfermion masses from the light messengers. This gives a constraint
\begin{equation}
\frac{M_{\rm mess}}{M_{\rm heavy}} \sim \epsilon^{\alpha_*/2}.
\end{equation}
Note that the contribution to sfermion masses from heavy messengers are suppressed by $m_{\rm mess}/M_{\rm heavy}$ relative to that from light messengers. Finally, we take the bino mass to be around $10^2$ GeV, so
\begin{equation}
\epsilon^{\gamma_X + 2} \frac{g_1^2}{16 \pi^2} M_{\rm mess} \sim 10^2 \; {\rm GeV}.
\end{equation}
All the conditions are satisfied, for example, if we take
\begin{equation}
M_{\rm mess} \sim 10^{11}\; {\rm GeV}, \quad M_{\rm heavy} \sim 10^{13}\; {\rm GeV}, \quad 
\sqrt{F} \sim 10^9\; {\rm GeV}, \nonumber
\end{equation}
\begin{equation}
\gamma_X \sim 1, \quad \alpha_* \sim 2.  \nonumber
\end{equation}
With these parameters, the gravitino mass is given by
\begin{equation}
m_{3/2} \sim {\mathcal O} (10^{-1}\;{\rm GeV}). \nonumber
\end{equation}
Again, the gravity and anomaly mediation effects are suppressed.

\subsection{$\mu/B_\mu$ Problem}

$\mu$ and $B_\mu$ parameters can be defined as coefficients of supersymmetric and non-supersymmetric holomorphic Higgs bilinear terms in the Lagrangian, respectively:
\begin{equation}
\int d^2 \theta \mu H_u H_d, \quad B_{\mu} H_u H_d.
\end{equation}
For natural electroweak symmetry breaking, both $\mu$ and $B_\mu$ should be typically in the same order as electroweak scale. $B_\mu$ is a supersymmetry breaking parameter, so should be in the same order as electroweak symmetry breaking scale if one assumes that hierarchy problem is solved by supersymmetry. However, $\mu$ is a supersymmetry preserving parameter and there is a priori no reason why it is also in the same order. This is so-called $\mu$ problem, and requires direct coupling between Higgs and hidden sector. 

In gauge mediated supersymmetry breaking, there is another problem called $B_\mu$ problem. Generally in models of gauge mediation, $\mu$ and $B_\mu$ are generated at the same loop order. For a model given in Appendix A, for example, both of them are generated at one-loop. Typically, 
\begin{equation}
\mu = \frac{\lambda_u \lambda_d}{16 \pi^2} \frac{F^\dagger}{M_{\rm mess}}, \quad
B_\mu = \frac{\lambda_u \lambda_d}{16 \pi^2} \frac{|F|^2}{M_{\rm mess}^2},
\end{equation}
where $\lambda_{u,d}$ is a coupling between $H_{u,d}$ and messengers and should be small for perturbation theory to be valid. This implies
\begin{equation}
\mu^2 \sim \frac{\lambda_u \lambda_d}{16 \pi^2} B_\mu \ll B_{\mu},
\end{equation}
which prevents natural electroweak symmetry breaking. For a general study and earlier attempts on $\mu$/$B_{\mu}$ problem, see \cite{muGGM} and references therein.

For our scenario, the gaugino masses are typically smaller than $\mu$, so we must set them to be in the same order by sequestering or small coupling constants $\lambda_{u,d}$. In this sense, $\mu$ problem is not solved completely naturally. Nevertheless, $B_{\mu}$ can be naturally taken in the same order as $\mu$ and other soft masses after conformal sequestering, following the line of arguments employed in \cite{SeqII,SeqIII}.

For concreteness, we work on a model where the next to leading order gaugino masses come from $F^3$-terms at one-loop, while the above argument also works in other models discussed in Section 3.1. Also, we consider all the soft parameters related to the Higgs sector for completeness. Among the soft terms, the $\mu$, and $A_{u,d}$ terms arise from operators linear in $X$. They are typically generated at one-loop by
\begin{equation}
c_{\mu} \int d^4 \theta X^\dagger H_u H_d, \quad c_{A_{u,d}} \int d^4 \theta X^\dagger H_{u,d}^\dagger H_{u,d},  
\end{equation} 
Here, $c_{\mu}$ and $c_{A_{u,d}}$ are numerical coefficients and proportional to $\lambda_u \lambda_d$ and $\lambda_{u,d}^2$, respectively. The scaling of these operators under hidden sector renormalization group flow is characterized by $\gamma_X$, the anomalous dimension of $X$. To have $\mu$ comparable to the gaugino masses, we should set $\lambda_u$ and $\lambda_d$ to satisfy 
\begin{equation}
\lambda_u \lambda_d \sim g^2_{\rm SM} \epsilon^{3 \gamma_X + \gamma_* + 4}.
\label{mucondition}
\end{equation}
This condition may require small $\lambda_u$ and $\lambda_d$, while that can be achieved if, e.g., the Higgs or messengers are composite.

The terms generated by operators quadratic in $X$ require a bit of care. The operators relevant to the Higgs sector are
\begin{equation}
c_{m_{u,d}} \int d^4 \theta XX^\dagger H_{u,d} H_{u,d}^\dagger, \quad c_{B_\mu} \int d^4 \theta XX^\dagger H_u H_d,
\end{equation}
The numerical coefficient are, up to order one coefficient, typically $(g_{\rm SM}^2/ 16 \pi^2)^2$, $\lambda_u \lambda_d/ 16 \pi^2$, respectively. At first sight, one may expect that $c_{m_{u,d}}$ and $c_{B_\mu}$ are suppressed by hidden sector renormalization group effects. As was pointed out in \cite{SeqII,SeqIII}, however, the operators quadratic in $X$ mix with those linear in $X$ and what ends up being suppressed is a specific combination of the coefficients $c_{m_{u,d}}$, $c_{B_\mu}$, $c_\mu$ and $c_{A_{u,d}}$. A careful operator product analysis \cite{Craig:2009rk} shows that the combinations that are sequestered by the anomalous dimension of $XX^\dagger$ are
\begin{equation}
c_{m_{u,d}} - \frac{1}{2} C (2-\alpha_*) (|c_\mu|^2 + |c_{A_{u,d}}|^2), \quad
c_{B_\mu} - \frac{1}{2} C (2 - \alpha_*) {\rm Re}(c_\mu (c_{A_u}^* +c_{A_d}^*)), \nonumber 
\end{equation}
where $C$ is the OPE coefficient of $X$ and $X^\dagger$ and assumed to be order one. Since $m_{u,d}^2 = - (c_{m_{u,d}} - |c_{A_{u,d}}|^2) |F|^2$ and $B_\mu = - (c_{B_\mu} - {\rm Re} (c_\mu(c_{A_u}^* + c_{A_d}^* ))) |F|^2$, they are naturally in the same order as $\mu$ and $A_{u,d}$ after sequestering if $\alpha_*$ is sufficiently large and $\lambda_u$ and $\lambda_d$ are taken to be in the same order. The value of $\alpha_*$ is constrained by
\begin{equation}
\frac{\lambda_u \lambda_d}{16 \pi^2} \gsim \epsilon^{\alpha_*}, \nonumber
\end{equation}
which can be rewritten, using \eqref{mucondition}, as 
\begin{equation}
\frac{g_{\rm SM}^2}{16 \pi^2} \gsim \epsilon^{\alpha_* - 3 \gamma_X - \gamma_* -4}. \nonumber
\end{equation}
This is satisfied for the choice of parameters in Section 3.2.

\section*{Acknowledgments}

We would like to thank M. Ibe, R. Kitano, Z. Komargodski, Y. Nakai and D. Shih for useful discussions. KH is grateful to the Perimeter Institute for Theoretical Physics for their hospitality during the course of this work. YO would like to thank the California Institute of Technology and the Kavli Institute for Theoretical Physics for their hospitality. KH's work was supported in part by the US Department of Energy under grant DE-FG02-95ER40899. YO's research at the Perimeter Institute for Theoretical Physics is supported in part by the Government of Canada through NSERC and by the Province of
Ontario through MRI. YO's research at KITP was supported in part by the National Science Foundation under Grant No. NSF PHY05-51164.


\appendix
\setcounter{equation}{0}
\renewcommand{\theequation}{A.\arabic{equation}}
\section*{Appendix A \, Generating $\mu$ and $B_\mu$}

In this Appendix, we present a simple model that generates $\mu$ and $B_\mu$ at leading order. Let us consider a simple messenger superpotential,
\begin{eqnarray}
W_{\rm mess} = m_1 (X) (\phi_1 \bar{\phi}_1 + \phi_2 \bar{\phi}_2) 
+ m_2 (X) (\phi'_1 \bar{\phi}'_1 + \phi'_2 \bar{\phi}'_2), \nonumber 
\end{eqnarray}
with coupling to Higgs being given by
\begin{eqnarray}
W_{\rm Higgs} = \lambda_u H_u \phi_1 \phi_2 + \lambda_d H_d\bar{\phi}_1 \bar{\phi}_2 
+  \lambda'_u H_u \phi'_1 \phi'_2 + \lambda'_d H_d\bar{\phi}'_1 \bar{\phi}'_2,  \nonumber 
\end{eqnarray}
The condition for anomalously small gaugino is given by 
\begin{eqnarray}
\frac{\partial}{\partial X} (m_1(X) m_2(X)) = 0.
\label{AnomalousCondition}
\end{eqnarray}
For this superpotential, the $\mu$-term at the leading order was calculated in \cite{DGP}:
\begin{eqnarray}
\mu &=& - \frac{F}{16 \pi^2} \left[ \lambda_u \lambda_d \left.\frac{m_1'(X)}{m_1(X)}\right|_0
+ \lambda'_u \lambda'_d \left.\frac{m_2'(X)}{m_2(X)}\right|_0\right]\nonumber\\
&=&  - \frac{F}{16 \pi^2} \left[ \lambda_u \lambda_d 
\left.\left(\frac{\partial}{\partial X} \ln m_1(X) m_2(X) \right)\right|_0
+ (\lambda'_u \lambda'_d - \lambda_u \lambda_d) \left.\frac{m_2'(X)}{m_2(X)}\right|_0\right]. \nonumber 
\end{eqnarray}
The first term in the second line vanishes because of \eqref{AnomalousCondition}, but the second 
term is non-zero in general (except for the case with $\lambda_u \lambda_d = \lambda'_u \lambda'_d$). 
We conclude that the $\mu$-term is generically generated at the leading order. Similarly, these superpotential interactions also generate $B_\mu$-term. The result is
\begin{equation}
B_\mu = - \frac{F}{16 \pi^2} \left[ \lambda_u \lambda_d \left(\left.\frac{m_1'(X)}{m_1(X)}\right|_0\right)^2 + \lambda'_u \lambda'_d \left(\left.\frac{m_2'(X)}{m_2(X)}\right|_0 \right)^2\right], \nonumber 
\end{equation}
which is also non-vanishing. In general, $\mu$ and $B_\mu$ do not vanish if each messenger couples to Higgs superfields differently.

%
%

\end{document}